\documentclass{sig-alternate}
\usepackage{hyperref}
\usepackage{xcolor}
\usepackage{footnote}

\begin{document}

\conferenceinfo{ADAPT'15}{, Jan 21 2015, Amsterdam, Netherlands.}

\title{A Self-adaptive Auto-scaling Method for Scientific Applications on HPC Environments and Clouds}
\subtitle{
}

%
%
%
%

\numberofauthors{3} 
%
\author{
%
%
\alignauthor
Kiran Mantripragada\\
       \affaddr{IBM Research - Brazil}\\
       \affaddr{S\~ao Paulo, Brazil}\\
       \email{kiran@br.ibm.com}
\alignauthor
Al\'ecio P. D. Binotto\\
       \affaddr{IBM Research - Brazil}\\
       \affaddr{S\~ao Paulo, Brazil}\\
       \email{abinotto@br.ibm.com}
\alignauthor
Leonardo P. Tizzei\\
       \affaddr{IBM Research - Brazil}\\
       \affaddr{S\~ao Paulo, Brazil}\\
       \email{ltizzei@br.ibm.com}
}

\maketitle
\begin{abstract}
High intensive computation applications can usually take days to months to finish an execution. During this time, it is common to have variations of the available resources when considering that such hardware is usually shared among a plurality of researchers/departments within an organization. On the other hand, High Performance Clusters can take advantage of Cloud Computing bursting techniques for the execution of applications together with on-premise resources. In order to meet deadlines, high intensive computational applications can use the Cloud to boost their performance when they are data and task parallel.
This article presents an ongoing work towards the use of extended resources of an HPC execution platform together with Cloud. We propose an unified view of such heterogeneous environments and a method that monitors, predicts the application execution time, and dynamically shifts part of the domain -- previously running in local HPC hardware -- to be computed in the Cloud, meeting then a specific deadline. The method is exemplified along with a seismic application that, at runtime, adapts itself to move part of the processing to the Cloud (in a movement called bursting) and also auto-scales (the moved part) over cloud nodes. Our preliminary results show that there is an expected overhead for performing this movement and for synchronizing results, but the outcomes demonstrate it is an important feature for meeting deadlines in the case an on-premise cluster is overloaded or cannot provide the capacity needed for a particular project.
\end{abstract}

\category{Computer systems organization}{Architectures}{Distributed architectures}[Cloud computing]
\category{Theory of computation}{De\-sign and analysis of algorithms}{Parallel algorithms}[Self-organization].



\section{Introduction}
\label{sec:intro}

High Performance Computing (HPC) platforms are designed to be explored by intensive computing applications to obtain performance, such as seismic analysis (for the exploration of oil and gas and natural disasters), weather and flooding forecasts, computational fluid dynamics, DNA sequencing, simulations of electromagnetic equation, among others. They are dedicated to provide their maximum throughput; and both applications and resources are usually fine-tuned to perform as optimally as possible for their specific challenges. Companies, government laboratories, and research/educational institutions acquire such HPC resources built on clusters of supercomputers, which are typically expensive to acquire and maintain. However, there is a considerable amount of time that such environments can stay idle, while other times it can be overloaded or even has capacity limitations that, together, prevents a particular project or simulation deadline to be achieved.

Cloud computing, on the other hand, proposes a resource-shared model in which users allocate resources on demand from providers only when necessary in a ``pay-as-you-run'' model, spending budget only for utilized processing time. Cloud was initially created to serve mostly Web applications deployed on Virtual Machines (VM) that share the same physical hardware with other VMs at the same time. As it matured, this model was coined Public Cloud and another model has emerged to deal with private data and also HPC requirements: the Private Cloud. It focuses on hosting computing intensive industrial and scientific workloads that deal with sensitive and large amounts of data -- on demand, but in a private and specialized space. This latter model have then specific requirements that are challenging to be met by Cloud providers, like the use of specialized resources (e.g., accelerators, GPUs, Infiniband, parallel file systems, etc), transfer and storage of large data sets, visualization, security and privacy, control,
 and a collaboration infrastructure.
Currently, providers are starting to offer Private Clouds with such specialized resources and this movement can eventually allows a hybrid execution, i.e., the use of on-premise HPC Clusters together with Clouds for boosting the execution platform at a certain period and to allow particular projects to meet their deadlines.

We propose a method to partition tasks and data of an application in both environments in an unified view. We use a seismic application as use case. This particular application can reflect the characteristics of a broad range of scientific and industrial applications: it is computational demand, data and task parallel, communication bound, utilizes solvers for differential equations, produces large amounts of data to be visualized as post-processing, etc. Our approach is able to monitor the application at runtime and to predict its total execution time so that it can perform a self-adaptation towards a Cloud Bursting, i.e., automatically shifts part of the data from a Cluster to be processed by the Cloud, in an unified and synchronized way, with the goal to meet a time deadline.

This article is organized as the following: Section~\ref{sec:approach} describes the proposed approach by emphasizing how the method performs the self-adaptation, auto-scaling, and automatic bursting, i.e., the application adapts its execution on both platforms at runtime; Section~\ref{sec:study} exemplifies the approach using a seismic application as an example and shows preliminary outcomes with a discussion; and Section~\ref{sec:rw} finalizes the proposal with related work, followed by conclusions on Section~\ref{sec:conclusion}.

\section{Dynamic and Self-Adaptive Cloud Bursting}
\label{sec:approach}

Let us consider an application that is data and task parallel, and which also evolves over time. Initially, the user starts the application according to specific parameters: \emph{data} (domain size), \emph{tasks}, number of \emph{timesteps}, \emph{partitions} of the data to be executed on-premise and in the cloud (data can be initialized fully placed on-premise). The system immediately starts to monitor the execution time for each timestep. The simulation continues to execute until the system detects that the \emph{threshold} time, i.e., the \emph{deadline} will not be satisfied. This situation can happen for a number of reasons, like concurrency in the local cluster, nodes down, dynamic change of the deadline, less on-premise capacity for a given problem, etc. Once the "\emph{time monitor}" detects a change in the estimated execution time, the system verifies this new estimation against the deadline time provided by the user. It is important to note that initial sequential timesteps are monitored in
order to reason whether they are predictable. In most cases, these industrial and scientific applications are simulations that evolves over time by solving a number of partial differential equations represented as matrices operations. Analyzing these characteristics, the timesteps can have similar execution times, but working on distinct data and, thus, being fairly predictable.

\begin{figure}[!htb]
	\centering
	\includegraphics[scale=0.35]{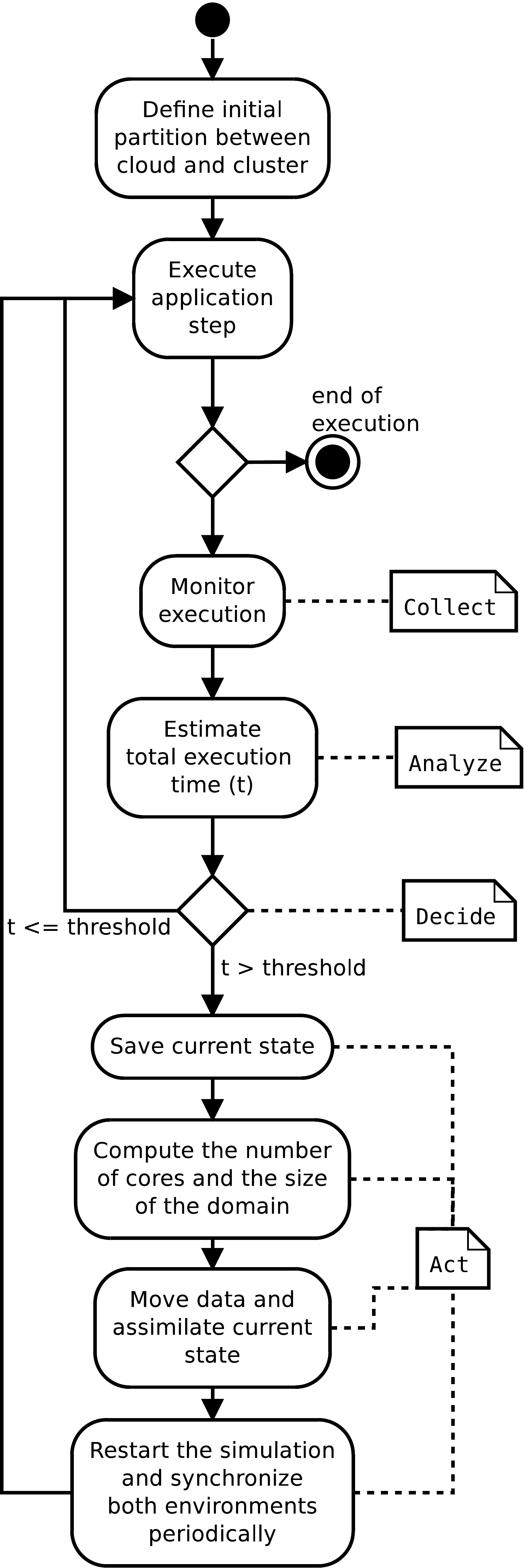}
	\caption{Steps of the self-adaptive method}
	\label{fig:method}
\end{figure}

If the monitoring module detects that the system will not attend the time threshold, this estimated time is used to compute the number of cores it would be needed to fit the simulation time within the deadline. This is done using a pre-processing phase to estimate the behavior of the application or it could also be an input parameter given by the user. Then, the system automatically starts the re-partitioning phase to migrate part of the domain to be computed by the elastic Cloud platform. Figure \ref{fig:method} and the following steps depict the proposed solution:

\begin{enumerate}
    \item Monitoring the application and analyzing estimated execution time to decide about a Cloud bursting;
	\item If a bursting is needed, save current state;
	\item Compute the number of cores to be allocated in the Cloud extended environment;
	\item Compute the size of domain (part of the  full domain) to be placed in the Cloud;
	\item Move current state data to new nodes (Cloud-burst);
	\item Assimilate current state as initial conditions;
	\item Restart simulation at the stopped step with new configuration (on-premise cluster and Cloud);
	\item Synchronize the simulation at each timestep running in both environments and merge results.
\end{enumerate}

In order to compute the estimated deadline and for validation purposes, we empirically defined the computational behavior of the application. By executing it for several cores and nodes configuration, a function that can represent the overall behavior was estimated:

\begin{equation}
	L_{cloud}(c) = - A \ln{c} + B
	\label{eq:logT1}
\end{equation}

\begin{equation}
	L_{cluster}(c) = -D \ln{c} + E.
	\label{eq:logT2}
\end{equation}

Equations \ref{eq:logT1} and \ref{eq:logT2} represent the application response as a function of number of cores, respectively for the Cloud environment and for an on-premise Cluster platform. The coefficients A, B, D and E were empirically computed for the experimental application using a small test job (small data set and few timesteps) as pre-processing---see Section \ref{sec:study}. The number of cores needed to attend the time threshold would satisfy only the Cluster environment, but since there is a possibility to place part of the domain to be executed in a Cloud environment, one must consider the network delays for migrating the data and restart the application as the reduced computing performance inherent in a Cloud environment when compared
to the cluster. The correction factor for the difference between the performances (Cluster x Cloud) can be further defined as follows:

\begin{align*}	
	K &= \frac{L_{cloud}}{L_{cluster}} \\
	K &= \frac{-A \ln c + B}{-D \ln c + E} \\
\end{align*}
	where $K$ is the computational performance correction factor and A, B, C and D are coefficients computed for a given application (Section \ref{sec:study}).
	
	Once the system computed the new number of cores required to attend the deadline, the correction factor can be applied:

\begin{equation}
	c_n = (c  - c_{cluster} ) * K
	\label{eq:newc}
\end{equation}
	where $c_n$ is the number of cores in the extended environment, $c$ is the estimated number of cores (Equation \ref{eq:logT2}) and $c_{cluster}$ is number of cores available in the on-premise cluster.

\begin{figure}[!htb]
	\centering
	\includegraphics[width=1\linewidth, height=0.7\linewidth]{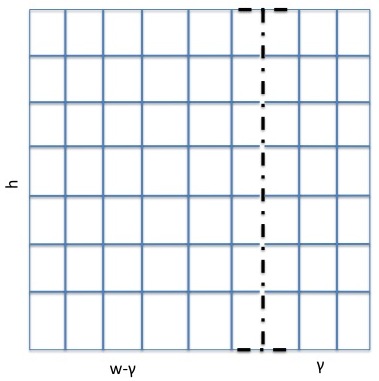}
	\caption{Domain split for Cloud-Bursting ($\gamma$ is the size to be placed in a Cloud)}
	\label{fig:split-domain}
\end{figure}

The next step is to split the domain into partitions to be executed in the hybrid environment.
In order to simplify the method, one of the two dimensions in domain (width and height) was fixed.
As depicted in  Figure \ref{fig:split-domain}, we chose to fix the height dimension to compute the value of $\gamma$ as a function of execution time.
That approach allows to define a linear relationship between the execution time (t) and the domain size through the value of $\gamma$:

\begin{equation}
	f(\gamma) = t = a\gamma + b
	\label{eq:gamma1}
\end{equation}
	where the linear coefficient \textbf{$a$} and the offset \textbf{$b$} must be computed for each application. \textbf{$\gamma$} must be any integer value, since it represents the number of column partitions to be placed in the extended environment.
	
	The Equation \ref{eq:gamma1} can be written as a function of time ($t$):
\begin{equation}
	 f^{-1} = g(t) = \gamma = \frac{t-b}{a}.
	\label{eq:gamma2}
\end{equation}
	The difference between estimated total time and the time for deadline is applied in Equation \ref{eq:gamma2} to compute the size of the partial domain $(\gamma)$ to be executed by the Cloud environment.

\section{Empirical Study}
\label{sec:study}

The goal of this study is to analyze the approach previously described for the purpose of creating a proof-of-concept model based on the application characteristics. Such analysis was carried out from the perspective of the developers of cloud bursting scientific applications targeting a seismic application. Obviously, the study involves two computing environments: on-premise Cluster environment and the Cloud computing environment. Both environments have different hardware (\textit{e.g.}, memory, CPU) and software configurations (\textit{e.g.}, operating system). The cloud environment is the \textit{IBM SoftLayer platform}\footnote{\url{http://www.softlayer.com/}}. Table~\ref{tab:hw} shows the hardware configuration of both environments. We have picked the cloud node configuration offer that was the most similar to the one in our local cluster for more correct evaluations.

\setlength{\tabcolsep}{1pt}
\renewcommand{\arraystretch}{1.5}
\begin{table}
	\caption{Hardware configuration of the Cluster and Cloud environments}
	\label{tab:hw}
	\begin{center}
	\footnotesize
	\begin{tabular}{c|c|c}
	          & \textbf{Cluster} & \textbf{Cloud} \\ \hline\hline
	processor (GHz) & 2.80 & 2.60 \\ \hline
	cores per processor & 10 & 4\\ \hline
	cache size (KB) & 25600  & 20480 \\ \hline
	total memory (KB) & 132127868 & 65711672 \\ \hline
	network & ethernet/infiniband & ethernet\\ \hline
	\end{tabular}
	\end{center}
\end{table}

\subsection{Target application: FWI}
\label{sec:targetApp}

Full Waveform Inversion (FWI) is a numerically challenging technique based on full wavefield modeling of a geological domain that extracts relevant quantitative information from seismograms~\cite{Virieux:2009:OFW}. When dealing with realistic physics-based elastic partial differential equations formulation and accurate discretization techniques, such as high order Spectral Element Method, the forward modeling becomes particularly challenging from the computational point of view. This type of modeling often requires grid representation of the order of billions of elements with interpolation polynomials of at least 4$^{th}$ degree in each spatial dimension, which entails hundreds of billions of variables computed for the forward solver. Furthermore, in the course of local optimization, the aforementioned process or its adjoint are repeated multiple times per non-linear inversion iteration. The computational complexity associated with FWI applications demands tremendous efforts to set up and maintain a 
distributed
scientific HPC infrastructure. These applications are also embedded in complex business models with various stakeholders. Therefore, having FWI applications offered as a service in the Cloud can bring several benefits to their users.

\begin{figure}[!htb]
	\centering
	\includegraphics[scale=0.4, trim=0cm 0cm 0cm 2cm, clip=true]{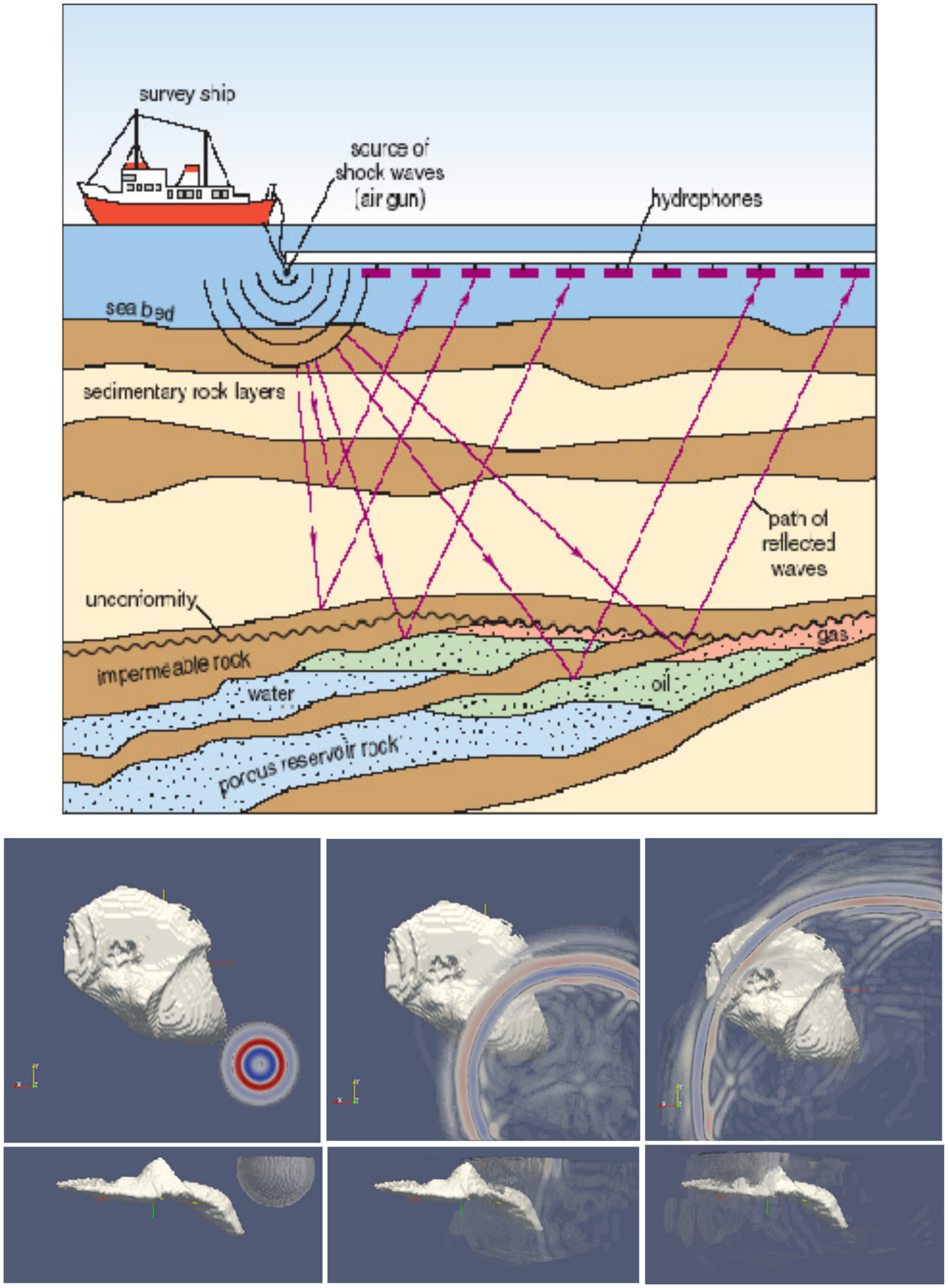}
	\caption{Examples of FWI data collection and remote visualization of the forward solver}
	\label{fig:fwi}
\end{figure}

Summarizing, millions of shots in different positions (i.e., independent tasks) are calculated over the same domain (i.e., data) and each shot solution is composed of several traces, evolving along timesteps. At the end, the generated images are composed to produce the final outcome. This process is depicted in Figure \ref{fig:fwi}. Above, multiple traces are produced by for shot by the acquisition ship that is moving over an area of exploration. The ship carries the compressed air gun along with an array of seismograms that capture wave reflections produced by different subsurface geological layers. Bellow, a produced timestep, in different point-of-views, is visualized. It represents a partial result of the reconstructed wave using our implementation of the FWI over a small region of interest of the area that contains a specific salt dome (we used the SEG/EAGE salt dome model).

This particular implementation (C++) is parallelized by means of OpenMPI (over data and tasks) and uses the Eigen3\footnote{\url{http://eigen.tuxfamily.org/}} library for matrix and vector manipulation, as well as ParaView\footnote{\url{http://www.paraview.org/}} to provide (remote) visualization. The application was executed several times in each environment and Table~\ref{tab:execution} presents the details of the executions.

\setlength{\tabcolsep}{1pt}
\renewcommand{\arraystretch}{1.5}
\begin{table}[!htb]
\caption{Execution details}
\label{tab:execution}
\begin{center}
\footnotesize
\begin{tabular}{c|c} \hline
number of elements (x axis)  & 600\\ \hline
number of elements (y axis)  & 600\\ \hline
polynomial order (x axis) & 4 \\ \hline
polynomial order (y axis) & 4 \\ \hline
degrees of freedom & 5764801\\ \hline
domain size (x axis) &  9000 \\ \hline
domain size (y axis) &  6000 \\ \hline
number of timesteps & 3000 \\ \hline
\end{tabular}
\end{center}
\end{table}

\subsection{Empirical Formulations}
\label{sub:empirical-formulations}

To validate our assumptions, we have performed some experiments on the target application.
The first equations were empirically determined through the results acquired from the simulations.
The chart on Figure~\ref{fig:timexcores} depicts that both curves have similar behavior, but as the number of processor-cores decreases to 10, the elapsed time for the FWI simulation   showed an increase of 150\% ($10^{5.4}/10^{5.0}$) worse than the on-premise cluster environment. On the other hand, as the number of processor-cores increases up to 40, the elapsed time of \textit{SoftLayer} environment was only around 50\% ($10^{4.3}/10^{4.1}$) longer than the cluster environment, tending to be the same with the increasing of cores. Furthermore, it is worth mentioning that these preliminary tests were executed with only a small number of cores and nodes to verify the behavior of Cloud, while running a high CPU and I/O intensive application.

From the Figure \ref{fig:timexcores}, we defined the equation \ref{eq:logT1} and \ref{eq:logT2} as follows:

\begin{figure}[!htb]
	\centering
	\includegraphics[width=1\linewidth]{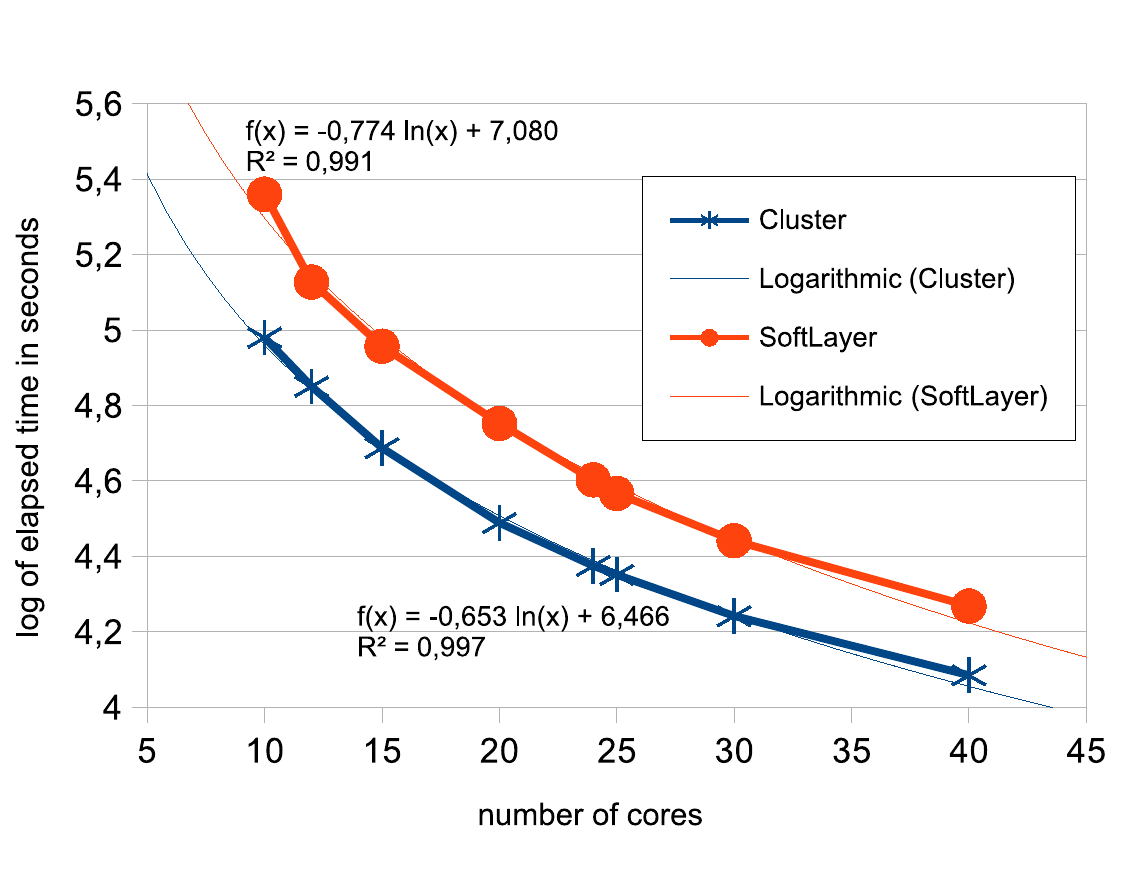}
	\caption{Execution time as a function of the number of processors}
	\label{fig:timexcores}
\end{figure}

\begin{equation}
	L_{cloud}(c) = - 0.77 \ln{c} + 7.1
	\label{eq:logT1_}
\end{equation}

\begin{equation}
	L_{cluster}(c) = -0.65 \ln{c} + 6.5
	\label{eq:logT2_}
\end{equation}

\begin{equation}
	 g(t) = \gamma = \frac{t - 231.18}{7.46}
	\label{eq:gamma2_}
\end{equation}

By solving equations \ref{eq:logT1_} to \ref{eq:gamma2_} for $c$, we can dynamically compute the value of $c_n$ in equation \ref{eq:newc} and \textbf{$\gamma$} in \ref{eq:gamma2_}. Those values are, respectively, the number of cores and the size of the domain placed in the Cloud environment. After applying the movement, the system is now capable to execute the simulation and satisfy the deadlines. This process is repeated while the system detects that the deadline time limit would not be attended for any reason, thus delaying the execution.

\subsection{Discussion}
\label{sub:discussion}

In this research, we propose a method that allows to dynamically move part of a domain to Cloud environments. In addition, the benefit of running the numerical solver in a Hybrid infrastructure (cluster and a private Cloud) showed interesting behavior when we forced the local HPC cluster to exhaust its limited resources.
Besides application performance gains, the proposed self-adaptive method reduces the time span needed for the "time to production" infrastructure, since there is no need to wait the acquisition, installation, and setup processes as would be with in acquiring new local resources, i.e., the co-existence in a balanced way of both platforms can bring better price-performance to such applications.

\begin{figure}[!htb]
	\centering
	\includegraphics[width=8.5cm, height=6.5cm]{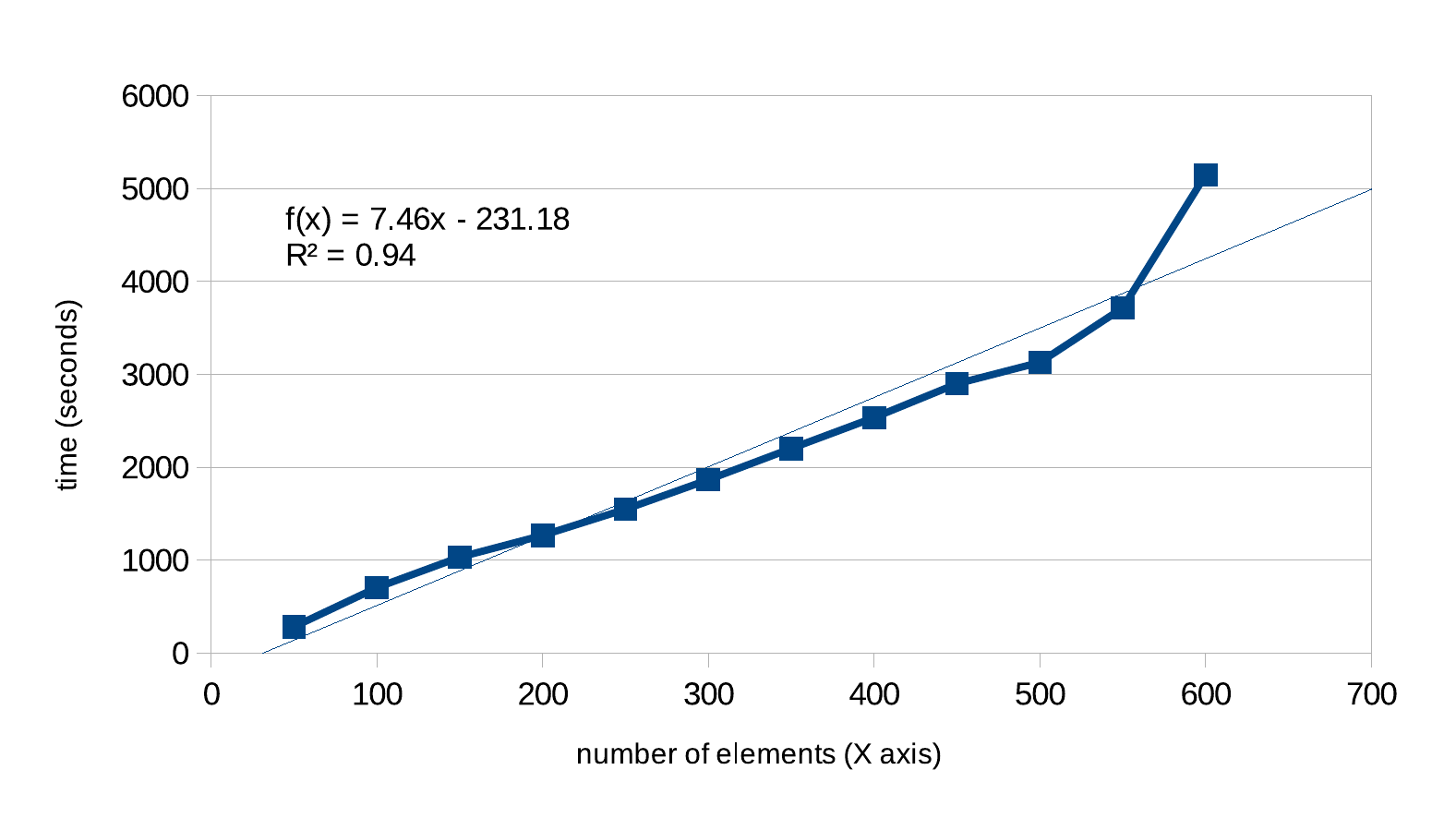}
	\caption{Empirical results for execution time (seconds) as a function of $\gamma$}
	\label{fig:timexelements}
\end{figure}

We performed the simulations over a few number of configurations and load-balancing strategies. The graph provided by Figure \ref{fig:timexelements} shows the application response for several grid sizes while holding one dimension constant.
Such an experiment allowed to understand how the domain split strategy would work when migrating part of the domain to an external environment. This study -- execution time when varying processor cores -- allowed to understand how to provision the Cloud cores for a given application size and deadline.
We then observed that the inverse process would also be an interesting way to understand how the execution time can be extrapolated to a specific number of cores. The self-adaptive strategy is based on the equations \ref{eq:logT1} to \ref{eq:logT2}, which allows to determine a few parameters to burst the application towards the Cloud. Those parameters, \textit{viz.} $\gamma$ and $c_n$, are sufficient to define the external infrastructure.

The overhead of the monitoring and partitioning components can be neglected. Results show that total message size is only $21KBytes$ size. The second-level partitioning strategy is based on stripes, which reduces the amount of communications between partitions and between environments.
It is important to note that the chosen load-balancing method over cores inside an environment is the simple greedy-based algorithm, where the striped-partitions are assigned to processor cores as long as they are available in each node. This choice reduces the inter-nodal communications as the neighbors partitions tend to be assigned to only one cluster node. In our case study, shots are independent tasks and traces have dependencies inside a shot.

On the other hand, the overhead caused by saving the actual state (checkpoint), the transferring corresponding data to the Cloud, and the provision the Cloud nodes need to be accounted. These values are not neglected and need to be considerable shorter than letting the application to finish its execution on the local cluster. Although we need to carefully analyze these measurements, in a seismic application -- which usually takes months to produce the final result -- such automatic \emph{checkpoint-restart} process is cost-effective as the total execution time of this application is inferred in the beginning of execution. We understand that this assumption would cause an inaccurate estimation and we are extending this work to measure and include such overhead as an offset value in equations \ref{eq:logT1} and \ref{eq:logT2}.

Despite of the potential feature for the system to be elastic, in the sense that infrastructure can expand and shrink as needed, the current approach just computes and applies expansions on the Cloud environment. Further studies and advances on this system can even consider to reduce the amount of processors when the deadline threshold can be achieved with reduced infrastructure.

Finally, the proposed method can be incorporated in a framework to be reused in different applications. For that, one needs to investigate the applications behavior before specifying the coefficient values as described in section \ref{sec:approach}. Considering that we simplified the mesh partitioning by fixing one of the dimensions, we are planing to extend this work by investigating the full 2D mesh partitioning and also full 3D numerical solvers.

\section{Related Work}
\label{sec:rw}

High Performance Computing applications are being tested on Cloud platforms. Works like \cite{gupta2013thewho}, \cite{gupta2011evaluation} and \cite{mateescu2011hybrid} performed a performance evaluation of a set of benchmarks and complex HPC applications on a range of platforms, from supercomputers to clusters, both in-house and in the cloud. These studies show that a Cloud can be effective for such applications mainly in the case of complementing supercomputers using models such as cloud burst and application-aware mapping to achieve significant cost benefits. Although these findings do not propose an automatic and adaptive approach for using both environments, their empirical studies opened the opportunity for proposals of tools that promote a hybrid approach based on these environments, like ours.

Analyzing Cloud as stand alone execution platform for HPC applications, like seismic, the authors of \cite{napper2009can} evaluated the Linpack workload on the Amazon EC2 cloud. Their conclusions indicate that the tested cloud environment has a potential, but it is not mature to provide a price-performance for HPC applications. \cite{ostermann2010performance} also evaluated the EC2 for a number of kernels used by HPC applications, coming also to a conclusion that such cloud services need an order of magnitude in performance improvement to better serve the scientific community. It is hard to evaluate one provider or another, but they are evolving to offers that are private and with specialized infrastructures, like with GPUs and Infiniband. In our study, we evaluated the \textit{SoftLayer} Cloud (virtualized nodes) and preliminary results indicate the environment as cost-effective in budget and performance when at least combined with on-premise clusters in a dynamic changing scenario.

More recently, \cite{belgacem2014hybrid} evaluated a computational fluid dynamics application over a heterogeneous environment of a cluster and the EC2 cloud. The results indicated that there is a need to adjust the CPU power (configuration) and workload by means of load-balancing. We are in line with this study and went further with the present work -- a dynamic self-adaptive method for application load-balancing over a hybrid platform composed of cluster and cloud.

\section{Conclusion}
\label{sec:conclusion}

We presented a first step towards a framework for self-adaptation of industrial and scientific applications in terms of being executed on a hybrid and heterogeneous environment composed of on-premise HPC clusters and the Cloud. It dynamically monitors and reasons when to migrate part of the computation from one platform to the other at runtime to maximize performance and meet deadline constraints. We demonstrated the core method applied to a seismic application, which is data and task parallel and could reflect the behavior of a broad range of industrial and scientific applications. The method is based on a self-adaptation of the application at runtime to shift part of the computation to the Cloud when the prediction of the total execution time is identified to overcome the given execution deadline (which could also change dynamically). The approach prevents additional expensive capital expenditure and has an intrinsic overhead, but due to the on-demand cloud elasticity of nodes
provisioning, the needed amount of nodes can be aggregated to achieve the deadline.

Ongoing work is based on the tool refinement, including overheads detection and shaping into a framework to be incorporated on the IBM's \textit{SoftLayer} Private and Public Clouds and/or even on the \emph{Bluemix platform}\footnote{\url{https://www.bluemix.net}} as a service targeting FWI as a service. Moreover, the roadmap includes to work with data from a real seismic exploration field acquisition and therefore a more robust cluster capable of processing the problem. Also, we intend to proceed with the analysis of nodes composed by CPU and GPU, thus including another level of heterogeneity inside the execution nodes (i.e., dynamically scheduling over the processors, like CPU and GPU, as reported in \cite{binotto:2013:balancing}).

Finally, the approach presented here at first sight might seem like a counter-intuitive idea: replace a straightforward embarrassingly parallel implementation of a challenging real world problem with a more sophisticated, self-adaptive, auto-scaled, synchronized, communication intense, domain partitioned approach. However, we have shown that a seamless combination of on-premise Clusters with the Cloud has an important contribution to boost the performance of computing intensive HPC applications with better price-performance and time-to-production.


\section{Acknowledgments}
This work has been partially supported by FINEP/MCTI under grant no. 03.14.0062.00.

%

\bibliographystyle{abbrv}

\end{document}